\documentclass[12pt]{iopart}
\usepackage{graphicx}
\usepackage{hyperref}
\expandafter\let\csname equation*\endcsname\relax
\expandafter\let\csname endequation*\endcsname\relax
\usepackage{amsmath}
\usepackage{mathtools}
\DeclarePairedDelimiter\abs{\lvert}{\rvert}%
\DeclarePairedDelimiter\norm{\lVert}{\rVert}%
\usepackage{placeins}

\let\Oldsection\section
\renewcommand{\section}{\FloatBarrier\Oldsection}

\let\Oldsubsection\subsection
\renewcommand{\subsection}{\FloatBarrier\Oldsubsection}

\let\Oldsubsubsection\subsubsection
\renewcommand{\subsubsection}{\FloatBarrier\Oldsubsubsection}

\makeatletter
\let\oldabs\abs
\def\abs{\@ifstar{\oldabs}{\oldabs*}}
\let\oldnorm\norm
\def\norm{\@ifstar{\oldnorm}{\oldnorm*}}
\makeatother

\makeatletter
\newcommand*{\centerfloat}{%
  \parindent \z@
  \leftskip \z@ \@plus 1fil \@minus \textwidth
  \rightskip\leftskip
  \parfillskip \z@skip}
\makeatother

\newcount\colveccount
\newcommand*\colvec[1]{
        \global\colveccount#1
        \begin{pmatrix}
        \colvecnext
}
\def\colvecnext#1{
        #1
        \global\advance\colveccount-1
        \ifnum\colveccount>0
                \\
                \expandafter\colvecnext
        \else
                \end{pmatrix}
        \fi
}

\begin{document}

\title[CTQW amplitudes and invariants]{Continuous-time quantum walks over simply connected graphs, amplitudes and invariants}

\author{Phillip R. Dukes}

\address{University of Texas Rio Grande Valley, Brownsville, TX 78520, USA}
\address{2 November 2015}
\ead{phillip.dukes@utrgv.edu}

\begin{abstract}
We examine the time dependent amplitude $ \phi_{j}\left( t\right)$ at each vertex $j$ of a continuous-time quantum walk on the cycle $C_{n}$. In many cases the Lissajous curve of the real vs. imaginary parts of each $ \phi_{j}\left( t\right)$ reveals interesting shapes of the space of time-accessible amplitudes. We find two invariants of continuous-time quantum walks. First, considering the rate at which each amplitude evolves in time we find the quantity $T = \displaystyle\sum_{j=0}^{n-1} \abs{\dfrac{d \phi_{j}\left( t\right)}{d t}}^{2}$ is time invariant. The value of $T$ for any initial state can be minimized with respect to a global phase factor $e^{i \theta t}$ to some value $T_{min}$. An operator for $T_{min}$ is defined. For any simply connected graph $g$ the highest possible value of $T_{min}$ with respect to the initial state is found to be $T_{min}^{max}=\left( \frac{\lambda_{max}}{2}\right)^{2}$ where $\lambda_{max}$ is the maximum eigenvalue in the Laplace spectrum of $g$. A second invariant is found in the time-dependent probability distribution $P_{j}\left(t\right) = \abs{\phi_{j}\left(t\right)}^{2}$ of any initial state satisfying $T_{min}^{max}$, with these conditions $\displaystyle\sum_{j=0}^{n-1}\left(P_{j}^{max} - P_{j}^{min}\right)^{2} = \dfrac{4}{n}$ for all simply connected graphs of $n$ vertices.
\end{abstract}

\section{Introduction}
Quantum walks on graphs come in two different versions, continuous-time and discrete-time. For a comprehensive review of quantum walks see Venegas-Andraca \cite{Venegas:2}. The main interests in quantum walks lie with the development of quantum search algorithms \cite{Ambainis:1, Ambainis:2, Ambainis:3}. In this paper, we concentrate on the amplitudes of a continuous-time quantum walk (CTQW) on connected graphs. A graph $G = (V,E)$ is defined by the set of vertices $V$ which are connected by the set of edges $E$. The adjacency matrix $A$ and degree matrix $D$ which describe the graph are defined as follows:
\begin{subequations}
\begin{align}
A_{j,k} &= \begin{cases} 1 &\mbox{if there is an edge connecting vertices \textit{j} and \textit{k}} \\
0 & \mbox{otherwise} \end{cases} \label{eq:Amat}\\
D_{j,k} &= \begin{cases} deg\left( v_{j}\right)  &\mbox{if }j=k \\
0 & \mbox{otherwise} \end{cases} \label{eq:Dmat}
\end{align}
\end{subequations}
where $deg\left( v_{j}\right)$ is the degree of vertex $v_{j}$.

 Generally, the unitary CTQW operator $U(t)$ is defined in terms of the Laplacian matrix $L$ of the graph \cite{Farhi:1, Gerhardt:1}
\begin{subequations}
\begin{align}
L &= D - A \label{eq:Laplacian}\\
U\left( t \right) &=e^{-i t L}. \label{eq:Uop}
\end{align}
\end{subequations}
For k-regular graphs, the operator is conventionally defined in terms of the normalized adjacency matrix $\frac{1}{k} A$ alone \footnote{This essentially amounts to a change in time-scale and a time-dependent global phase factor.} \cite{Ahmadi:1, Ben-Avraham:1}
\begin{equation}
U\left(t\right)  = e^{-i t \frac{1}{k} A}.
\label{eq:Ureg}
\end{equation}
The quantum walker is described by a time-dependent quantum state $\lvert \phi\left( t\right)  \rangle$. The initial state vector $\lvert \phi\left( 0\right)  \rangle$ is an ordered set of components, each component corresponding to the initial amplitude at each vertex of the graph (we will label each vertex starting with $0$)

\begin{equation}
\lvert \phi\left( 0\right)  \rangle  =\left[\phi_{0}\left( 0\right) ,\phi_{1}\left( 0\right),\phi_{2}\left( 0\right),...,\phi_{n-1}\left( 0\right)\right] 
\label{eq:initialState}
\end{equation}
such that
\begin{equation}
\langle\phi\left( 0\right)  \lvert \phi\left( 0\right)  \rangle  =  \displaystyle\sum_{j=0}^{n-1} \abs{\phi_{j}\left( 0\right)}^2 = 1 .
\label{eq:initialNorm}
\end{equation}
The time development of the quantum walk then becomes
\begin{equation}
U\left(t\right)\lvert \phi\left(0\right)  \rangle  = \lvert \phi\left(t\right)  \rangle,
\label{eq:QW}
\end{equation}
and the time-dependent amplitude at vertex $j$ is 
\begin{equation}
 \phi_{j}\left( t\right)  = \langle j \lvert \phi\left( t\right)  \rangle.
\label{eq:amp-at-j}
\end{equation}

\section{The time-dependent amplitudes at each vertex of $C_{n}$}
\begin{figure}[!htbp]
\centering
\includegraphics[scale=0.55]{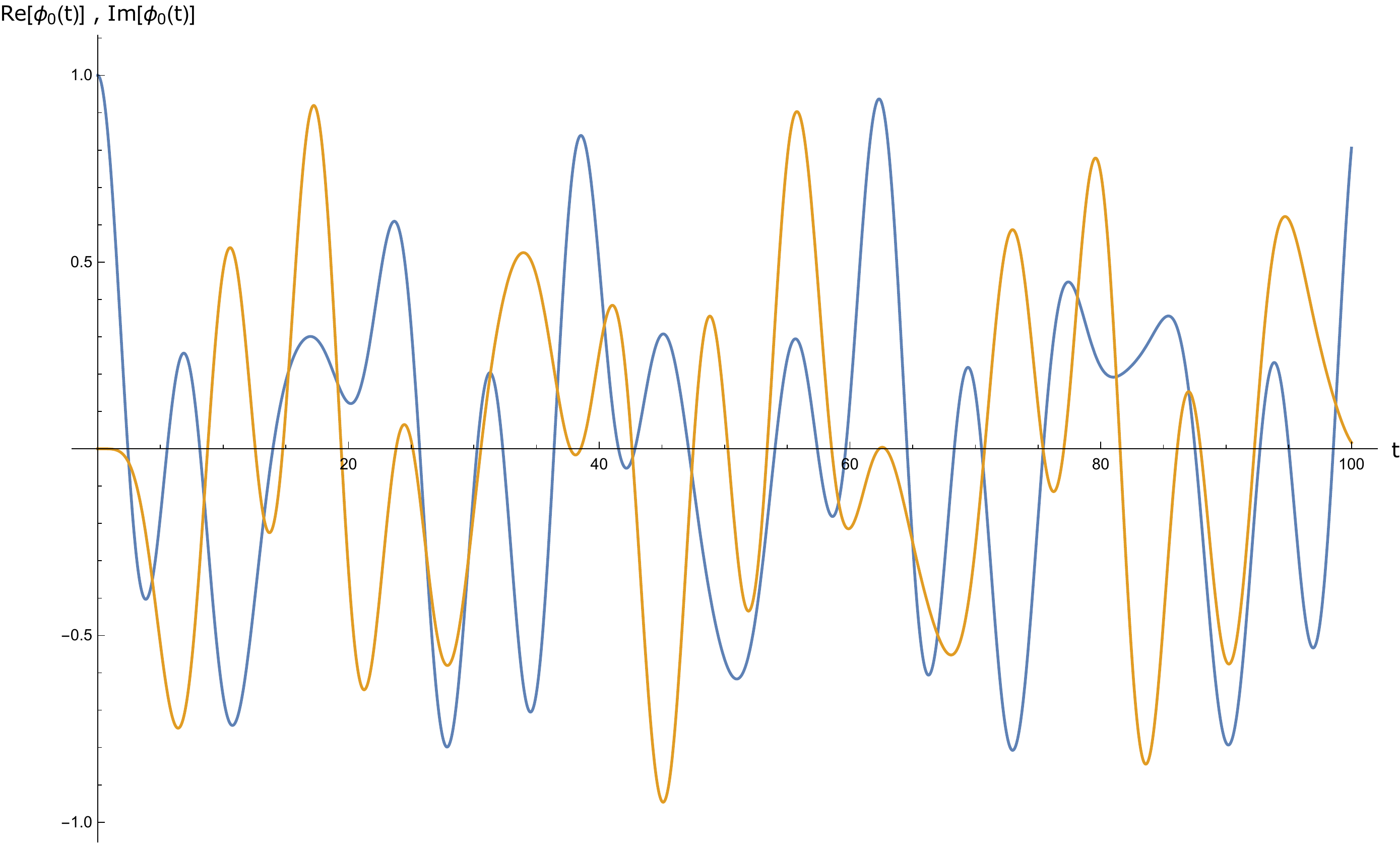}
\caption{A plot of the time-dependent real and imaginary parts of $\phi_{0}\left(t\right)$ on a cycle $C_{5}$.  }
\label{fig:C50}
\end{figure}

\begin{figure}[!htbp]
\centerfloat
\includegraphics[scale=1.0]{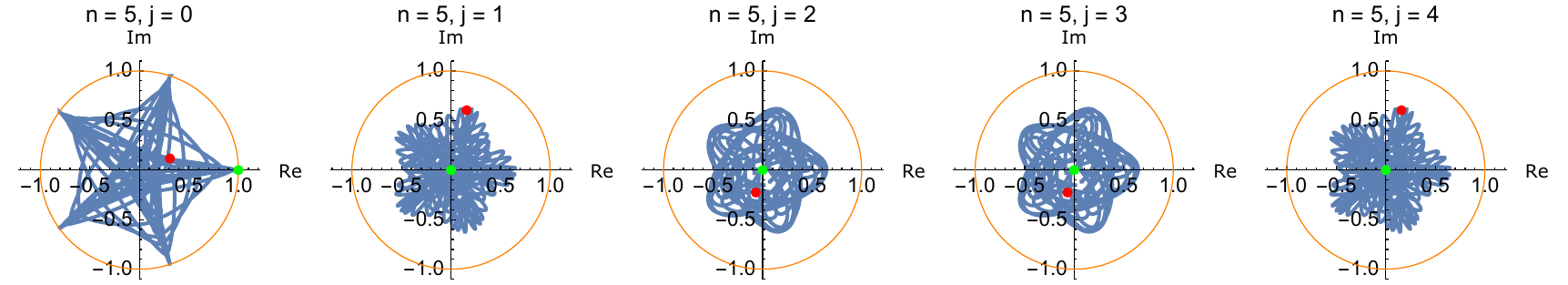}
\caption{The Lissajous curves of the time-dependent amplitude at each vertex of $C_{5}$ when the quantum walker is initially localized at vertex $j=0$, plotted from $0\leq t \leq 200$. (The unit circles are included for scale. The green and red dots indicate the initial and final values in each Lissajous curve.)}
\label{fig:C5}
\end{figure}

\begin{figure}[!htbp]
\centering
\includegraphics[scale=0.55]{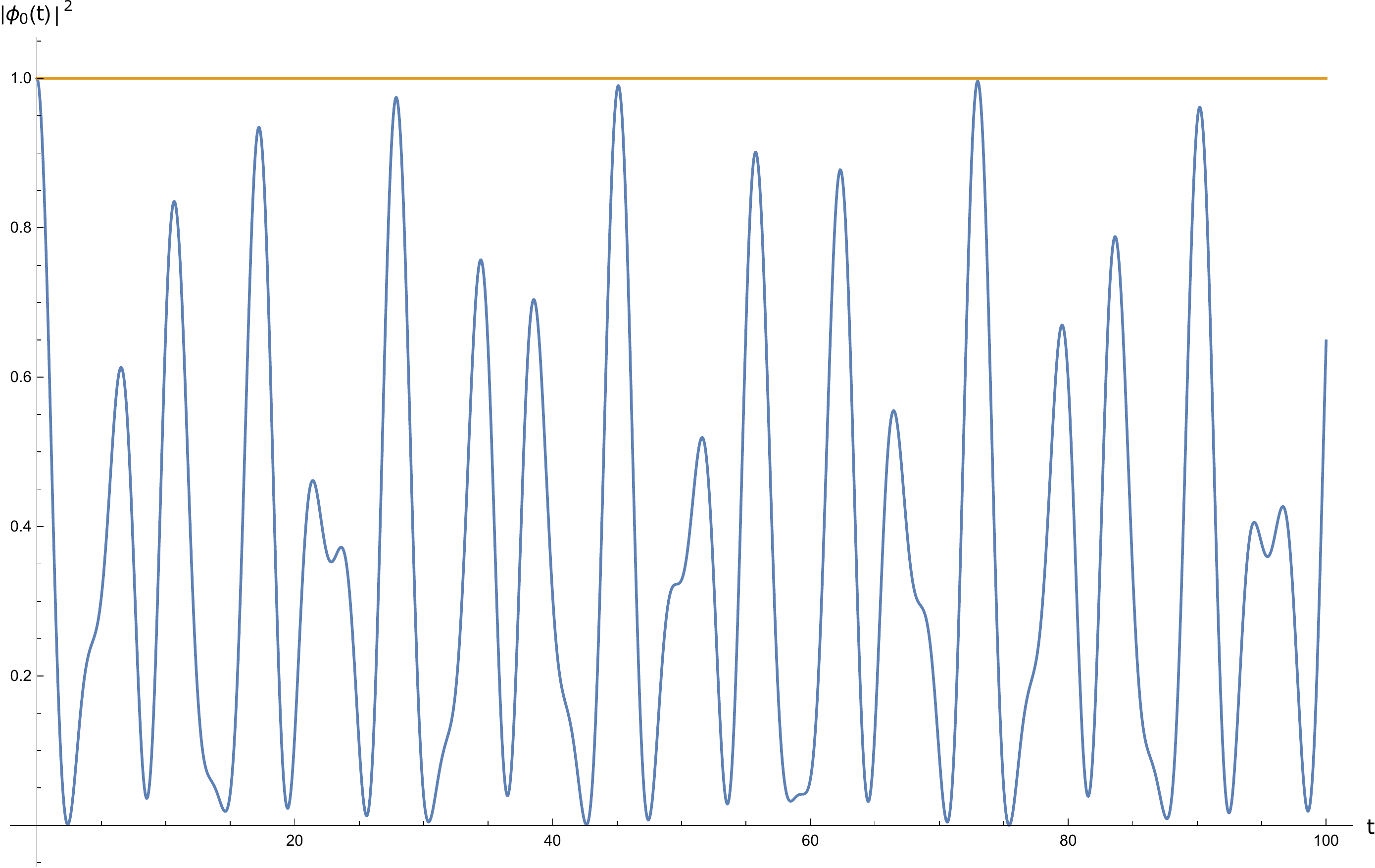}
\caption{The plot of $\vert \phi_{0}\left( t\right) \vert^{2}$ corresponding to Fig. \ref{fig:C5} from $0\leq t \leq 100$. $\vert \phi_{0}\left( t\right) \vert^{2} \simeq 0.996259$ at $t\simeq 72.9715$.}
\label{fig:C5a}
\end{figure}

A cycle graph $C_{n}$ is a 2-regular graph with $n$ vertices. The corresponding $U\left(t\right)$ will be an $n \times n$ circulant matrix completely defined by an $n$ component circulant vector  $\mathbf{\alpha}\left(t\right)$\cite{Wyn-jones:1}
\begin{equation}
U\left(t\right)  = CIRC_{n}\left[  \alpha_{0}\left( t\right) ,\alpha_{1}\left( t\right),\alpha_{2}\left( t\right),   \:\cdot\,\cdot\,\cdot\:, \alpha_{n-1}\left( t\right) \right].
\label{eq:CIRC}
\end{equation}
For any cycle $C_{n}$ the components $ \alpha_{j}\left( t\right) $ are given by Ben-Avraham, et al. \cite{Ben-Avraham:1} as
\begin{equation}
 \alpha_{j}\left( t\right)  = \frac{1}{n} \displaystyle\sum_{k=0}^{n-1}e^{- i t \cos\left( 2\pi \frac{k}{n}\right)} e^{2\pi i \frac{j k}{n}}. 
\label{eq:ampsCn}
\end{equation}
This is equivalent to the definition in equation \ref{eq:Ureg}. The amplitudes $\alpha_{j}\left( t\right)$ are periodic only for $n\in\left\lbrace 1,2,3,4,6\right\rbrace$. This is due to the fact that all the $\cos \left( 2\pi \frac{k}{n}\right) $ terms in equation \ref{eq:ampsCn} are rational only when $n\in\left\lbrace 1,2,3,4,6\right\rbrace$. Thus, $C_{5}$ is the smallest cycle with aperiodic amplitudes. When the quantum walker is initially localized at vertex $j=0$, i.e., $\phi_{0}\left( 0\right)=1$, the time-evolved amplitude at each vertex is equal to the corresponding component of the circulant vector, $\phi_{j}\left( t\right)=\alpha_{j}\left(t\right)$. A plot of the real and imaginary parts of $\phi_{0}\left( t\right)$ is given in Fig. \ref{fig:C50}. There is no obvious phase relationship between the real and imaginary parts; however, the conventional tool for investigating phase relations between two different waveforms is the Lissajous curve or parametric plot.
The Lissajous curve of the real vs. imaginary parts of each amplitude is presented in Fig. \ref{fig:C5}.
Because the $\phi_{j}\left( t\right)$ are aperiodic, as $t \rightarrow \infty$, the Lissajous curves will become densely filled-in and the shape defines a continuous space of accessible amplitudes. 
A plot of the probability at vertex $j=0$,  $P_{0}\left(t\right)=\vert \phi_{0}\left( t\right) \vert^{2}$ is given in Fig. \ref{fig:C5a}. The greatest probability in the time interval $0<t\leq 100$ is $P_{0}\simeq 0.996259$ at $t\simeq72.9715$.

\begin{figure}[!htbp]
\centering
\includegraphics[scale=1.0]{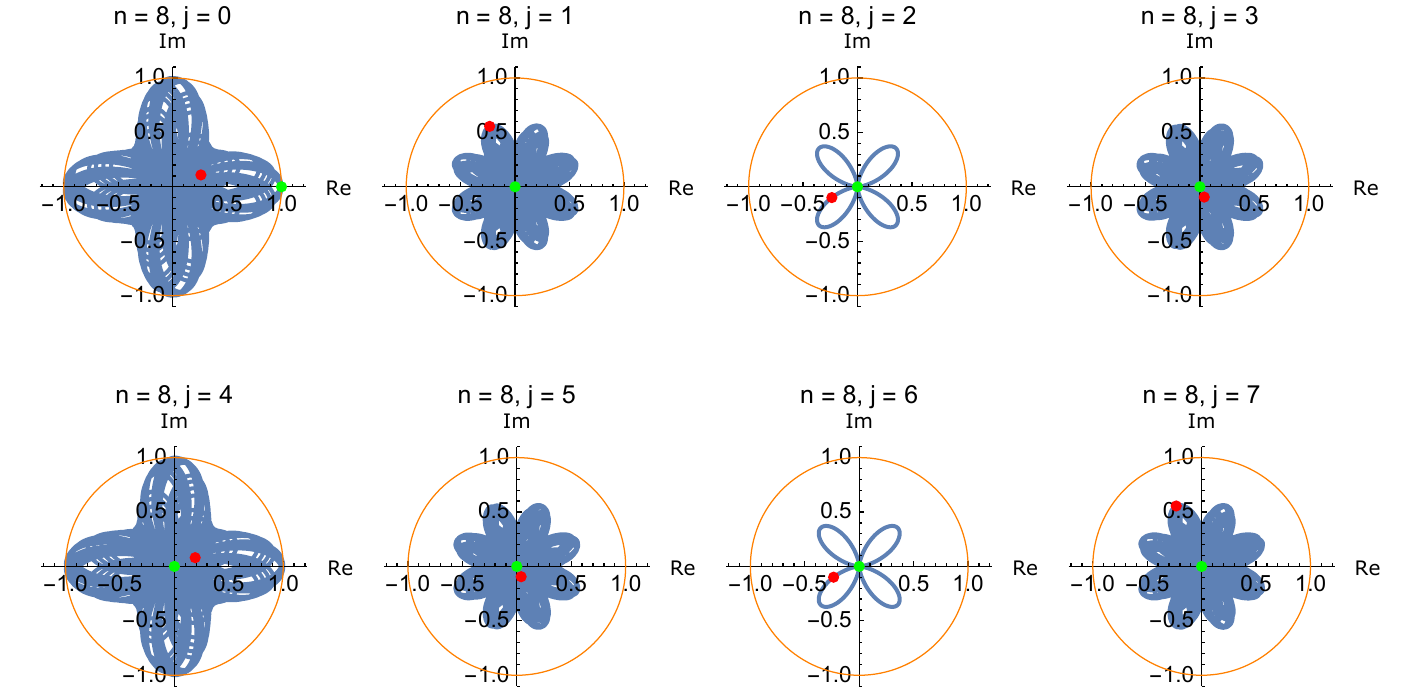}
\caption{The Lissajous curves of a CTQW on $C_{8}$. The quantum walker is initially localized at vertex $j=0$. Note that the amplitudes at vertices $j=2,6$ are periodic. (Equation \ref{eq:ampsCn} is augmented with a time-dependent global phase factor of $e^{-i\frac{1}{4}t}$.)}
\label{fig:C8}
\end{figure}

\begin{figure}[!htbp]
\centering
\includegraphics[scale=0.55]{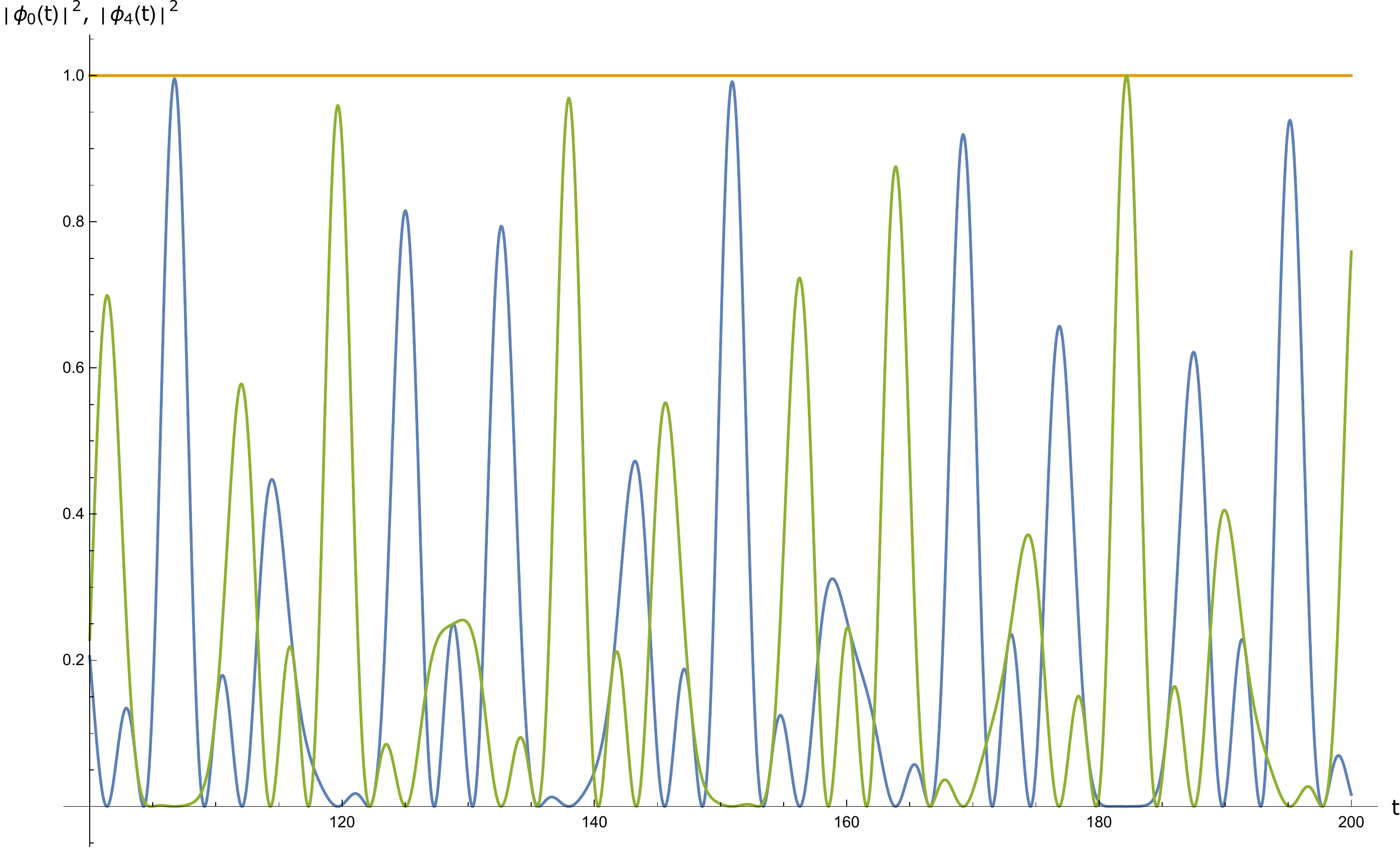}
\caption{Plots of $\vert \phi_{0}\left( t\right) \vert^{2}$ and $\vert \phi_{4}\left( t\right) \vert^{2}$ corresponding to Fig. \ref{fig:C8} from $100\leq t \leq 200$. $\vert \phi_{0}\left( t\right) \vert^{2}\simeq 0.995731$ at $t\simeq 106.722$ and $\vert \phi_{4}\left( t\right) \vert^{2}\simeq 0.999633$ at $t\simeq 182.185$.}
\label{fig:C8a}
\end{figure}

When $n\in \left\lbrace \mathtt{even\: integers} \right\rbrace $ equation \ref{eq:ampsCn} yields amplitudes which are purely real or purely imaginary. In these cases, to create Lissajous curves with descriptive phase relationships we introduce a time-dependent global phase factor. There is a subjective component to what the ``right" phase factor should be. By trial and error we find that a global phase factor of $\alpha(t) = e^{-i \frac{1}{4} t}$ has an evocative effect. The Lissajous curves for $C_{8}$ with initial state $\phi_{0}\left(0\right)=1$ is presented in Fig. \ref{fig:C8}. Plots of $\vert \phi_{0}\left( t\right) \vert^{2}$ and $\vert \phi_{4}\left( t\right) \vert^{2}$ are given in Fig. \ref{fig:C8a}.

Inspection of Figs. \ref{fig:C5} and \ref{fig:C5a} suggest that the quantum walker will nearly return to its initial state at an aperiodic rate. Likewise, inspection of Figs. \ref{fig:C8} and \ref{fig:C8a} suggests the quantum walker will essentially oscillate (though aperiodically) between vertices $j=0$ and $j=4$.  It is an open question if there exists a time $t>0$ such that $\vert \phi_{0}\left( t\right) \vert ^{2} = 1$ when $\vert \phi_{0}\left( 0\right) \vert ^{2} = 1$ for any $C_{n}$ when $n \notin \left\lbrace 1,2,3,4,6\right\rbrace$. The problem of recurrence of the initial state of the quantum walker in a discrete quantum walk on $C_{n}$ has been solved by Dukes \cite{Dukes:1}.

We have inspected the amplitudes of a CTQW on $C_{n}$ through $n=201$. With increasing $n$ the amplitudes become distributed over a greater number of vertices. Consequently, the area covered by the Lissajous curves become progressively smaller as $n$ increases, also the shapes are often surprising.

\section{A kinematic invariant of a CTQW}
The rate at which the quantum walker state vector $ \lvert \phi_{\left( t\right)}\rangle$ changes in time is
\begin{equation}
\begin{aligned} 
\lvert \dot{\phi}_{\left( t\right)}\rangle & = \frac{d}{dt} \lvert \phi_{\left( t\right)}\rangle \\
&=  \frac{d}{dt}U\left(t\right)\lvert \phi_{\left(0\right)}\rangle \\
&= -i L e^{-i t L}\lvert \phi_{\left(0\right)}\rangle.
\end{aligned}
\label{eq:Vamp}
\end{equation}
Because an operator commutes with a function of itself and L is real-Hermitian we arrive at
\begin{equation}
\begin{aligned} 
\langle \dot{\phi}_{\left( t\right)} \lvert \dot{\phi}_{\left( t\right)}  \rangle & =  \displaystyle\sum_{i=0}^{n-1} \abs{ \dfrac{d \phi_{i}\left( t\right) }{d t}}^{2} \\
&=  \langle\phi_{\left( 0\right)}\rvert L^{2} \lvert\phi_{\left( 0\right)} \rangle \\
&= T.
\end{aligned}
\label{eq:Vamp2}
\end{equation}
It is tempting to think of $T$ as the ``total kinetic energy" of the CTQW on a graph. The value of $T$ is not invariant to a time-dependent global phase factor. Taking $\lvert \phi^{\prime}_{\left(t\right)}\rangle = e^{i \theta t}\lvert \phi_{\left(t\right)}\rangle$ we readily obtain
\begin{equation}
\begin{aligned} 
\langle \dot{\phi}^{\prime}_{\left( t\right)} \lvert \dot{\phi}^{\prime}_{\left( t\right)}\rangle & = \langle \phi_{\left( 0\right)} \rvert (L-\theta)^{2} \lvert \phi_{\left( 0\right)} \rangle \\
&= T\left(\theta \right).
\end{aligned}
\label{eq:Vamptheta}
\end{equation}

\begin{figure}[!htbp]
\centering
\includegraphics[scale=0.55]{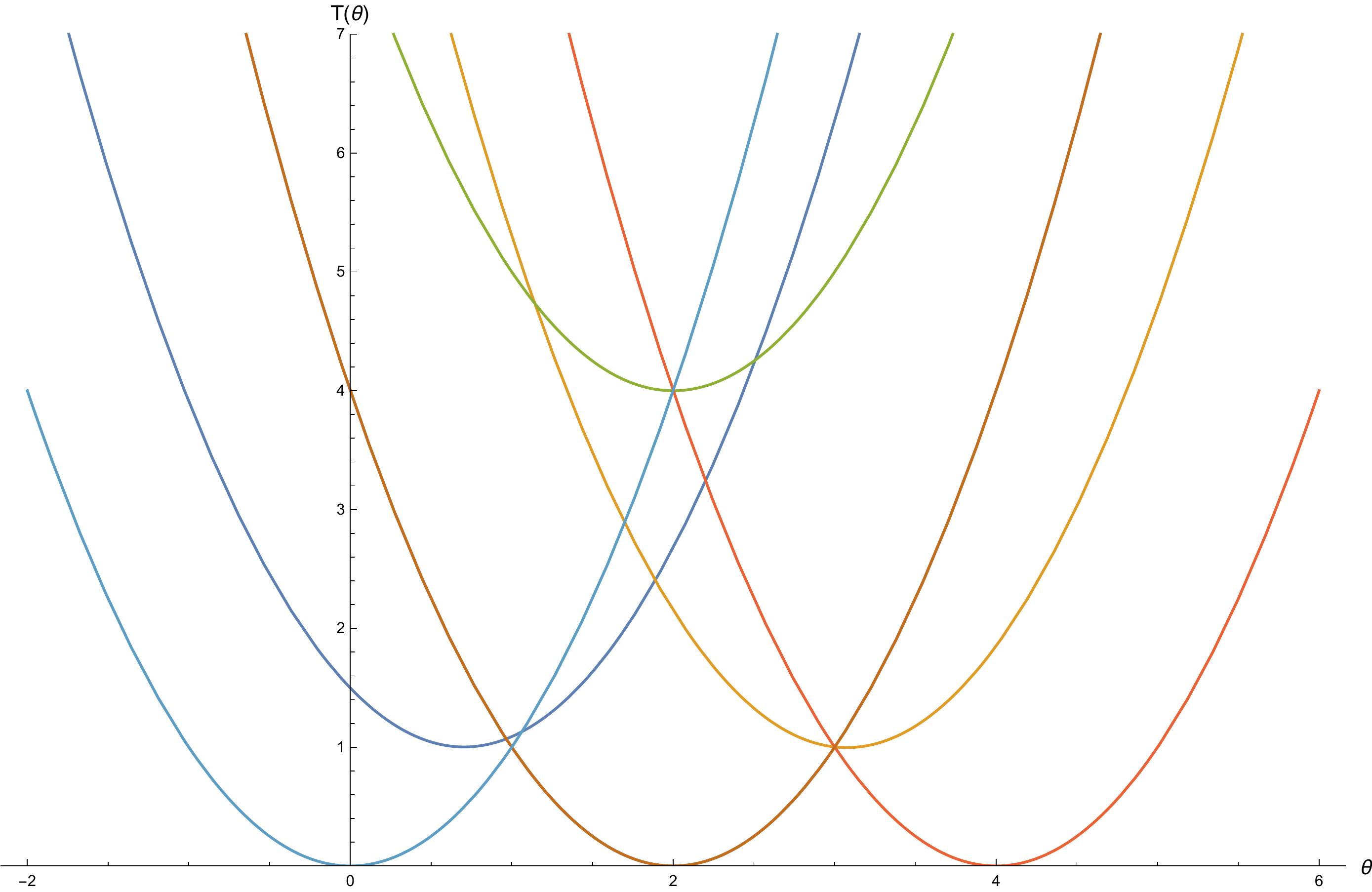}
\caption{Plots of $T\left(\theta\right)$ for 6 different initial states on the cycle graph $C_{4}$.}
\label{fig:Ttheta}
\end{figure}

The Hermitian matrix $L$ has a full set of orthonormal eigenvectors $\{\lvert v_{0}\rangle, \lvert v_{1}\rangle,...,\lvert v_{n-1}\rangle\}$ with corresponding eigenvalues $\{\lambda_{0}, \lambda_{1},...,\lambda_{n-1}\}$. The initial state $\lvert \phi_{\left(0\right)}\rangle$ can be expressed as a superposition of the eigenvectors
\begin{equation}
\lvert \phi_{\left(0\right)}\rangle = a_{0}\vert v_{0}\rangle + a_{1}\vert v_{1}\rangle + \cdot\cdot\cdot + a_{n-1}\vert v_{n-1}\rangle.
\label{eq:initalV}
\end{equation}
In this representation $T\left(\theta\right)$ becomes
\begin{equation}
\begin{aligned} 
T\left(\theta\right)&= \langle\phi_{\left(0\right)}\vert\phi_{\left(0\right)}\rangle \theta^{2} - 2 \langle\phi_{\left(0\right)}\vert L\vert\phi_{\left(0\right)}\rangle \theta + \langle\phi_{\left(0\right)}\vert L^{2}\vert\phi_{\left(0\right)}\rangle\\
&= \left(\displaystyle\sum_{j=0}^{n-1}\abs{a_{j}}^{2} \right)\theta^{2} - 2\left(\displaystyle\sum_{j=0}^{n-1}\abs{a_{j}}^{2} \lambda_{j} \right)\theta + \left(\displaystyle\sum_{j=0}^{n-1}\abs{a_{j}}^{2} \lambda_{j}^{2} \right)\\
&= A \theta^{2} + B \theta + C.
\end{aligned}
\label{eq:Vamppara}
\end{equation}
$T\left(\theta\right)$ is a convex parabola with a vertical axis of symmetry. Plots of $T\left(\theta\right)$ for a few selected initial states on $C_{4}$ are shown in Fig. \ref{fig:Ttheta}.
For any initial state there will be a minimum value $T_{min}=T\left(\theta_{v}\right)$ where $\theta_{v}$ is the abscissa value of the coordinates for the vertex of the parabola. A fundamental relation in geometry gives the coordinates of the vertex in terms of the coefficients of equation \ref{eq:Vamppara}
\begin{equation}
\begin{aligned} 
\left(\theta_{v},T_{min}\right)  &= \left( -\frac{B}{2A}, -\frac{B^{2}-4AC}{4A}\right) \\
&= \left(\langle\phi_{\left(0\right)}\vert L\vert\phi_{\left(0\right)}\rangle,\langle\phi_{\left(0\right)}\vert L^{2}\vert\phi_{\left(0\right)}\rangle - \langle\phi_{\left(0\right)}\vert L\vert\phi_{\left(0\right)}\rangle^{2}\right)
\end{aligned}
\label{eq:minVertex}
\end{equation}
where we have $A = \langle\phi_{\left(0\right)}\vert\phi_{\left(0\right)}\rangle = 1$. 

We find that $T_{min}$ can be represented as an operator $\mathrm{T_{min}}$ acting on the idempotent density matrix $\rho$ representing the initial state
\begin{equation}
\begin{aligned} 
\mathrm{T_{min}}\,\rho  &= L^{2}\rho -  L\rho L\rho \\
&= L^{2}\rho\rho -  L\rho L\rho \\
&= L\left(L\rho - \rho L\right) \rho \\
&= L\left[L,\rho\right]\, \rho,
\end{aligned}
\label{eq:TminOpp}
\end{equation}
so that $T_{min}= \mathrm{Tr}\left(L\left[L,\rho\right]\, \rho\right)$.

\section{The extrema of $T_{min}$ with respect to the initial state}
$T_{min}$ is a measure of the total ``dynamism" of the amplitude $\vert\phi_{\left(t\right)}\rangle$ minimized with respect to a global phase factor $e^{i \theta t}$. In other words, the dynamism is due to the time dependent norm of each component of $\vert\phi_{\left(t\right)}\rangle$. Thus, the stationary states of the operator $U(t)$ will have the lowest possible value $T_{min}=0$. These states correspond to the normalized eigenvectors of the Laplacian matrix $L$. A stationary state, as is indicative of the name, will have a constant, time-independent, norm and thus the probability distributions over the vertices of the graph will be constant in time. Initial states with values of $T_{min} >0$ will produce probability distributions which vary in time. Higher values of $T_{min}$ are due to the range in norm values as well as their rate of change. 

The maximization of $T_{min}$, $T^{max}_{min}$, does not define a unique initial state. The eigenvalues of the Laplacian matrix $L$ can be put in increasing order
\begin{equation}
0=\lambda_0 \leq \lambda_1 \leq \lambda_2 \leq \cdots \leq \lambda_{n-1}.
\label{eq:ordval}
\end{equation}
If the graph is simply connected, $\lambda_1 > 0$. If the greatest eigenvalue $\lambda_{n-1} = \lambda_{max}$ is degenerate with multiplicity $m$ then
\begin{equation}
\lambda_{max} = \lambda_{n-1} = \lambda_{n-2} = \cdots = \lambda_{n-m}.
\label{eq:degen}
\end{equation}
The initial states which maximize $T_{min}$ are a superposition of the eigenvectors associated with the eigenvalue $0$ and an arbitrary unit vector in the eigenspace of $\lambda_{max}$ of the form
\begin{equation}
\begin{aligned}
\lvert \phi_{\left(0\right)} \rangle &= \frac{1}{\sqrt{2}}\lvert v_0 \rangle + \frac{1}{\sqrt{2}}\left(\alpha_{1} \lvert v_{n-1}\rangle + \alpha_{2} \lvert v_{n-2}\rangle + \cdots + \alpha_{m} \lvert v_{n-m} \rangle  \right),\\
\lvert \alpha_1 \rvert^{2} &+ \lvert \alpha_2 \rvert^{2} + \cdots + \lvert \alpha_{m} \rvert^{2} = 1.
\end{aligned}
\label{eq:maxTphi}
\end{equation}
Then
\begin{equation}
T^{max}_{min} = \langle\phi_{\left(0\right)}\vert L^{2}\vert\phi_{\left(0\right)}\rangle - \langle\phi_{\left(0\right)}\vert L\vert\phi_{\left(0\right)}\rangle^{2} = \left( \frac{\lambda_{max}}{2}\right)^{2}.
\label{eq:Tmax}
\end{equation}

\section{An invariant in the probability distribution $\lvert \phi_{i}\left(t\right)\rvert^{2}$ of $T^{max}_{min}$ states}
The time development of the initial state defined in equation \ref{eq:maxTphi} will be
\begin{equation}
U(t)\lvert \phi_{(0)}\rangle = \frac{e^{-i\lambda_{0}t}}{\sqrt{2}}\lvert v_0 \rangle + \frac{e^{-i\lambda_{max}t}}{\sqrt{2}}\left(\alpha_{1} \lvert v_{n-1}\rangle + \alpha_{2} \lvert v_{n-2}\rangle + \cdots + \alpha_{m} \lvert v_{n-m} \rangle  \right)
\label{eq:maxTphit}
\end{equation}
This time dependent state can yield different time dependent probability distributions over the vertices of the graph according to different values for the amplitudes $\alpha_j$. Each $P_{j}\left(t\right)=\lvert \phi_{j}\left(t\right)\rvert^{2}$ will be either constant or periodic with a period of $\frac{2 \pi}{\lambda_{max}}$. An invariant among these distributions will be the sum of the square of the difference between the maximum and minimum of each $P_{j}\left(t\right)$,
\begin{equation}
\displaystyle\sum_{j=0}^{n-1}\left(P_{j}^{max} - P_{j}^{min}\right)^{2} = \dfrac{4}{n},
\label{eq:Pinv1}
\end{equation}
for any simply connected graph of $n$ vertices. The proof is as follows; For any simply connected graph with $n$ vertices the normalized eigenvector corresponding with the eigenvalue $\lambda_{0}=0$ will be of the form
\begin{equation}
\lvert v_{0}\rangle = \frac{1}{\sqrt{n}}\colvec{5}{1}{1}{\cdot}{\cdot}{1}.
\label{eq:v0}
\end{equation}
The arbitrary unit vector in the eigenspace of $\lambda_{max}$ can be represented as
\begin{equation}
\begin{aligned}
&\left(\alpha_{1} \lvert v_{n-1}\rangle + \alpha_{2} \lvert v_{n-2}\rangle + \cdots + \alpha_{m} \lvert v_{n-m} \rangle  \right)= \colvec{5}{a_{0}}{a_{1}}{\cdot}{\cdot}{a_{n-1}},\\
&\lvert a_{0}\rvert^2 + \lvert a_{1}\rvert^2 + \cdots + \lvert a_{n-1}\rvert^2 =1.
\end{aligned}
\label{eq:arbvec}
\end{equation}
Equation \ref{eq:maxTphit} can then be written as
\begin{equation}
\begin{aligned}
\lvert \phi\left(t\right)\rangle &=\frac{1}{\sqrt{2n}}\colvec{5}{1}{1}{\cdot}{\cdot}{1} + \frac{e^{-i \lambda_{max}t}}{\sqrt{2}} \colvec{5}{a_{0}}{a_{1}}{\cdot}{\cdot}{a_{n-1}}\\
\colvec{5}{\phi_{0}\left(t\right)}{\phi_{1}\left(t\right)}{\cdot}{\cdot}{\phi_{n-1}\left(t\right)}&=\frac{1}{\sqrt{2}}\colvec{5}{\frac{1}{\sqrt{n}}}{\frac{1}{\sqrt{n}}}{\cdot}{\cdot}{\frac{1}{\sqrt{n}}} + \frac{1}{\sqrt{2}} \colvec{5}{r_{0}e^{-i\left( \lambda_{max} t+ \theta_{0}\right)}}{r_{1}e^{-i\left( \lambda_{max} t+ \theta_{1}\right) }}{\cdot}{\cdot}{r_{n-1}e^{-i\left( \lambda_{max} t+ \theta_{n-1}\right) }} 
\end{aligned}
\end{equation}
where $r_{0}^2 + r_{1}^2 + \cdots + r_{n-1}^2 = 1$.

Each $\phi_{j}\left(t\right)$ will have its greatest or lowest norm when the corresponding $e^{-i\left(\lambda_{max}t+\theta_{j}\right)}$ term equals $+1$ or $-1$ respectively.
\begin{subequations}
\begin{align}
\phi^{+}_j &= \frac{1}{\sqrt{2n}}+\frac{r_{j}}{\sqrt{2}} \label{eq:phi+}\\
\phi^{-}_j &= \frac{1}{\sqrt{2n}}-\frac{r_{j}}{\sqrt{2}} \label{eq:phi-}
\end{align}
\end{subequations}
The maximum and minimum probability values $P^{max}_{j}$ and $P^{min}_{j}$ are then
\begin{subequations}
\begin{align}
P^{max}_{j}=\lvert\phi^{+}_j \rvert^2 &= \frac{1}{2n}+ \frac{r_{j}}{\sqrt{n}}+\frac{r^2_{j}}{2} \label{eq:P+}\\
P^{min}_{j}=\lvert\phi^{-}_j \rvert^2 &= \frac{1}{2n}- \frac{r_{j}}{\sqrt{n}}+\frac{r^2_{j}}{2} \label{eq:P-}
\end{align}
\end{subequations}
such that
\begin{equation}
\displaystyle\sum_{j=0}^{n-1}\left(P_{j}^{max} - P_{j}^{min}\right)^{2} = \dfrac{4}{n},
\label{eq:Pinv2}
\end{equation}



\section*{References}
\bibliographystyle{unsrt}
\bibliography{QWrefs}

\end{document}